\documentclass[journal]{IEEEtran}

\usepackage{siunitx}
\usepackage{booktabs}
\usepackage{hyperref}
\usepackage{soul}
\usepackage{caption}
\usepackage{cite}
\usepackage{amsmath,amssymb,amsfonts}
\usepackage{algorithmic}
\usepackage{graphicx}
\usepackage{textcomp}
\usepackage{seqsplit}
\usepackage{xcolor}
\usepackage{makecell}
\usepackage{enumitem}
\usepackage{babel}
\usepackage{hyperref}
\def\BibTeX{{\rm B\kern-.05em{\sc i\kern-.025em b}\kern-.08em
    T\kern-.1667em\lower.7ex\hbox{E}\kern-.125emX}}
    
\usepackage{amsthm}
\newtheoremstyle{cited}{3pt}{3pt}{\itshape}{}{\bfseries}{.}{.5em}{\thmname{#1} \thmnumber{#2}\thmnote{\normalfont#3}}

\theoremstyle{definition}

\theoremstyle{cited}

\usepackage{hyperref}

\begin{document}

\title{Review of the NIST Light-weight Cryptography Finalists}

\author{
    \IEEEauthorblockN{
    William J Buchanan\IEEEauthorrefmark{1}, Leandros Maglaras\IEEEauthorrefmark{1}}\\
    \IEEEauthorblockA{\IEEEauthorrefmark{1}Blockpass ID Lab, Edinburgh Napier University}}

\maketitle

\begin{abstract}
Since 2016, NIST has been assessing light-weight encryption methods, and, in 2022, NIST published the final 10: ASCON, Elephant, GIFT-COFB, Grain128-AEAD, ISAP, Photon-Beetle, Romulus, Sparkle, TinyJambu, and Xoodyak. The time
that the article was written, NIST announced ASCON as the chosen method that will be published as NIST’s lightweight
cryptography standard later in 2023. In this article we provide a comparison between these methods in
terms of energy efficiency, time for encryption and time for hashing.
    \end{abstract}
    

\section{Introduction}
While AES and SHA work well together within computer systems, they struggle in an IoT/embedded world as they take up: too much processing power; too much physical space; and consume too much battery power. So NIST outlines a number of methods which can be used for light-weight cryptography, and which could be useful in IoT and RFID devices \cite{buchanan2017lightweight}. They define the device spectrum as:

\begin{itemize}
\item    Conventional cryptography. Servers and Desktops. Tablets and smart phones.
\item    Light-weight cryptography. Embedded Systems. RFID and Sensor Networks.
\end{itemize}

With embedded systems, we commonly see 8-bit, 16-bit and 32-bit microcontrollers, and which would struggle to cope with real-time demands for conventional cryptography methods. And in the 40+ years since the first 4-bit processor, there is even a strong market for 4-bit processors. RFID and sensor network devices, especially, have limited numbers of gates available for security, and are often highly constrained with the power drain on the device.

So AES is typically a non-starter for many embedded devices. In light-weight cryptography, we often see smaller block size (typically 64 bits or 80 bits), smaller keys (often less than 90 bits) and less complex rounds (and where the S-boxes often just have 4-bits).

For light-weight cryptography the main constraints that we have are typically related to ener requirements, gate equivalents (GEs), and timing. With passive RFID devices, we do not have an associated battery for the power supply, and where the chip must power itself from energy coupled from the radio wave. An RFID device is thus likely to be severely constrained in the power drain associated with any cryptography functions, along with being constrained for the timing requirements and for the number of gates used. Even if an RFID device has an associated battery (active RFID), it may be difficult to recharge the battery, so the drain on power must often be minimised.

There is thus often a compromise between the cryptography method used and the overall security of the method. Thus often light-weight cryptography methods balance performance (throughput) against power drain and GE, and do not perform as well as main-stream cryptography standards (such as AES and SHA-256). Along with this, the method must also have a low requirement for RAM (where the method requires the usage of running memory to perform its operation) and ROM (where the method is stored on the device). In order to assess the strengths of various methods we often define the area that the cryptography function will use on the device – and which is defined in $\mu m^2$.

And, so, since 2016, researchers have been probing the submitted methods, and in 2022 NIST published the final 10: ASCON, Elephant, GIFT-COFB, Grain128-AEAD, ISAP, Photon-Beetle, Romulus, Sparkle, TinyJambu, and Xoodyak. A particular focus is on the security of the methods, along with their performance on low-cost FPGAs/embedded processes and their robustness against side-channel attacks. The time that the artile was written, NIST announced ASCON as the chosen method that will be published as NIST's lightweight cryptography standard later in 2023. Nevertheless in this article we provide a comparison between these methods in terms of energy efficiency, time for encryption and time for hashing using an Arduino Due with an ARM Cortex M3 running at 84MHz, Arduino Nano Every (AVR ARmega 4809), Arduino MKR Zero (ARM Cortex
M10+) and Arduino Nano 33 BLE.

\section{Related work}
\label{related_work}

Bellini et al \cite{bellini2022randomness} outlined a set of tests for the NIST finalists related to their randomness. These tests were similar to the tests performed for the AES and the SHA-3 standards. The tests include: Avalanche Plaintext; Avalanche Key; Plaintext-Ciphertext correlation; Cipher Block Chaining Mode; Random; Low-Density with Plaintext ;Low-Density with Key; High-Density with Plaintext; and High-Density with Key. The main conclusion is that the underlying primitives for most of the methods produce datasets that seem random.  in the first third of the total number of rounds. 

Elsadek et al \cite{elsadek2022hardware} performed an evaluation on the NIST finalists. In terms of the size of each implementation (Table \ref{tab:els}, we see that TinyJambu and Grain128 have the smallest footprint, while Sparkle has by far the largest footprint. In terms of gate equivalent (GE), TinyJambu requires 3,600 GEs, while Sparkle needs 39,500.

\begin{table*}

\caption{\label{tab:els} Average energy efficiency \cite{elsadek2022hardware}}
\centering
\begin{tabular}{|l|l|l|l|}
\hline
& Area ($\mu m^2$ - synthesis over 22nm)&Area (kGE)\\
\hline\hline
TinyJambu&716&3.6\\
Grain128-AEAD&861&4.3\\
GIFT-COFB&1,618&8.1\\
Romulus&1,961&9.8\\
ASCON&21,93&11\\
Xoodyak&2,387&11.9\\
Photon-Beetle&2,523&12.6\\
ISAP&3,080&15.4\\
Elephant&3,458&17.3\\
Sparkle&7,897&739.5\\
\hline
\end{tabular}
\end{table*}


\section{Methods}\label{sec:Background}
Table \ref{tab:lwc_summary} outlines some of the core specifications of the NIST finalists \cite{madushan2022review}.

\begin{table*}

\caption{\label{tab:lwc_summary} Outline specifications for NIST LWC finalist algorithms \cite{madushan2022review}}
\centering
\begin{tabular}{|l|l|l|l|l|l|l|l|l|l|}
\hline
Name &  Type &  Variant &  Underlying Primitive &  State &  Key  &  Mode  &  Rate/Block  &  Tag  &  Security \\
 &   &   &   &   (Bits) &   (Bits) &    &  (Bits) &   (Bits) &   (Bits)\\
\hline\hline
Ascon &  Sponge &  Ascon-128 &  Ascon-p &  320 &  128 &  Duplex &  64 &  128 &  128\\
&&          Ascon-128a &  Ascon-p &  320 &  128 &  Duplex &  128 &  128 &  128\\
\hline
Elephant &  Sponge &  Jumbo &  Spongent &  176 &  128 &  Elephant &  176 &  64 &  127\\
&& Dumbo &  Spongent &  160 &  128 &  Elephant &  160 &  64 &  112\\
&&         Delirium &  Keccak &  200 &  128 &  Elephant &  176 &  128 &  127\\
\hline
GIFT-COFB &  Block &  GIFT-COFB &  GIFT-128 &  192 &  128 &  COFB &  128 &  128 &  128\\
\hline
Grain-128AEAD &  Stream &  Grain-128AEAD &  N/A &  256 &  128 &  N/A &  1 &  64 &  128\\
\hline
ISAP &  Sponge &  ISAP-A-128 &  Ascon-p &  320 &  128 &  ISAP &  64 &  128 &  128\\
&&          ISAP-K-128 &  Keccak &  400 &  128 &  ISAP &  144 &  128 &  128\\
&&          ISAP-K-128A &  Keccak &  400 &  128 &  ISAP &  144 &  128 &  128\\
&&          ISAP-A-128A &  Ascon-p &  320 &  128 &  ISAP &  64 &  128 &  128\\
\hline
PHOTON-Beetle &  Sponge &  PHOTON-Beetle-AEAD[128] &  PHOTON256 &  256 &  128 &  Beetle &  128 &  256 &  121\\
&&         PHOTON-Beetle-AEAD &  PHOTON256 &  256 &  128 &  Beetle &  32 &  256 &  128\\
\hline
Romulus &  Block &  Romulus-M &  Skinny-128-384 &  384 &  128 &  COFB &  128 &  128 &  128\\
&&          Romulus-N &  Skinny-128-384 &  384 &  128 &  COFB &  128 &  128 &  128\\
&&          Romulus-T &  Skinny-128-384 &  384 &  128 &  COFB &  128 &  128 &  128\\
\hline
SPARKLE &  Sponge &  SCHWAEMM256-128 &  SPARKLE &  384 &  128 &  SPARKLE &  256 &  128 &  120\\
&&          SCHWAEMM128-128 &  SPARKLE &  256 &  128 &  SPARKLE &  128 &  128 &  120\\
&&          SCHWAEMM192-192 &  SPARKLE &  384 &  192 &  SPARKLE &  192 &  192 &  184\\
&&          SCHWAEMM256-256 &  SPARKLE &  512 &  256 &  SPARKLE &  256 &  256 &  248\\
\hline
TinyJambu &  Sponge &  TinyJambu &  TinyJambu &  128 &  128 &  TinyJambu &  32 &  64 &  120\\
\hline
Xoodyak &  Sponge &  Xoodyak &  Xoodoo &  384 &  128 &  Cyclist &  352 &  128 &  128\\
\hline

\end{tabular}
\end{table*}

\subsection{ASCON}
ASCON \cite{dobraunig2016ascon, ASCON} was designed by Christoph Dobraunig, Maria Eichlseder, Florian Mendel and Martin Schläffer from Graz University of Technology, Infineon Technologies, and Radboud University. It is both a lightweight hashing and encryption method. ASCON uses a single lightweight permutation with Sponge-based modes of operation and an SPN (substitution–permutation network) permutation. Overall it has an easy method of implementing within hardware (2.6 gate equivalents) and software. A 5-bit S-box (as used in Keccak’s S-box core) is used to enable a lightweight approach, and it has no known side-channel attacks. It can also achieve high throughputs such as throughputs of between 4.9 and 7.3 Gbps. It stores its current state with 320 bits.

\subsection{Elephant}
Elephant is a light-weight crypto cipher created by Tim Beyne, Yu Long Chen, Christoph Dobraunig, and Bart Mennink \cite{beyne2020dumbo}. It is an authenticated encryption scheme, based on a nonce-based encrypt-then-MAC construction. We can provide a nonce (or IV) and which is the salt value for the cipher. Along with this it supports Authenticated Encryption with Associated Data (AEAD) and where we can provide additional data for the cipher. This additional data is not sent with the cipher or used within the cipher, but can be used to authenticate it. An example might be to link a packet sequence number to the additional data, so that the cipher could not be played-back for another sequence number. One of the strengths of Elephant is that it has a low footprint such as with a 160-bit permutation and can be parallelized.

 Within Elephant we have three different variations: Dumbo, Jumbo and Delirium (and which is a Belgian beer with a pink elephant logo). It was created by Tim Beyne, Yu Long Chen, Christoph Dobraunig, and Bart Mennink \cite{Elephant}. Tim and Yu are from KU Leuven and imec-COSIC, Belgium, and Christoph and Bart from Radboud University, The Netherlands.

One of the strengths of Elephant is that it has a low footprint and can use a 160-bit permutation. This small value reduces the footprint of the method within memory. In order to speed up the encryption process, the method can also be parallelized. There are three main methods:

\begin{itemize}
\item Dumbo: Elephant-Spongent-$\pi$[160]. This method is well-matched to hardware and gives the baseline level of security. The 160 value relates to the bit size of the permutation that is operated on. Overall it gives 112-bit security levels.
\item Jumbo: Elephant-Spongent-$\pi$[176]. This is an improved method and achieves 127-bit security.
\item Delirium: Elephant-Keccak-f[200]. This is a more software-focused approach with reasonably good hardware performance, and with 127-bit security.
\end{itemize}

\subsection{GIFT-COFB}
GIFT-COFB is a light-weight crypto cipher created by Subhadeep Banik, Avik Chakraborti, Tetsu Iwata, Kazuhiko Minematsu Mridul Nandi, Thomas Peyrin, Yu Sasaki, Siang Meng Sim and Yosuke Todo \cite{banik5gift, GIFT, asecuritysite_73133}. It uses a COFB (COmbined FeedBack) block cipher based AEAD mode using the GIFT-128 block cipher \cite{banik2017gift}. Overall GIFT is similar to a smaller version of PRESENT. It is thought that GIFT has better performance than SIMON and SKINNY. GIFT-COFB fixes some of the security weaknesses of PRESENT. As many systems are now looking to authenticate encryption with AEAD (Authenticated Encryption with Additional Data). With AEAD we add in extra data and which can be used to authenticate the connected cipertext. For example, with network packets, we could add the source port and sequence number into the additional data, and where these were used to perform the authentication. The AEAD addition was then added through a submission to the NIST competition for lightweight cryptography with GIFT-COFB. 

\subsection{Grain128-AEAD}
Grain is a Light Weight Stream Cipher and was written by Martin Hell, Thomas Johansson and Willi Meier \cite{hell2006stream, asecuritysite_58074}. It has a relatively low gate count, power consumption and memory. In its AEAD form, it has a 128-bit key, and has two shift registers (LFSR - Linear Feedback Shift Register and NFSR - Non-linear Feedback Shift Register) and a nonlinear output function. The gate equivalent (GE) relates to the level of parallelization defined. For the lowest level of parallization (s=1), we can achieve a GE of 3,638, and with the maximum parallelization level (s=32), we have a GE of 12,110. When running a maximum speed for the lowest level of parallelization we get a throughput of 560 Mbps, a power drain of 3.6~mW and an area of 5,258 $\mu m^2$. With the highest level of parallelization we can achieve a throughput of 10.59~Gbps. 

\subsection{ISAP}
Isap is a lightweight block cipher and was written by Christoph Dobraunig, Maria Eichlseder, Stefan Mangard, Florian Mendel, Bart Mennink, Robert Primas and Thomas Unterluggauer \cite{dobraunig2020isap, isap, asecuritysite_12892}. It is focused on robustness against power analysis and fault attacks and where there is a node for small code size. Overall it uses a sponge-based mode with SPN permutations. Isap has mechanisms to protect against fault attacks. Isap uses an Encrypt-then-MAC design with two keys for IsapMac and IsapEnc. 

\subsection{Photon-Beetle}
One of the great advantages of using a sponge method in cryptography is that you get the addition of hashing on top of encryption for very little overhead in the code size and the memory requirements. If we are using something like an 8-bit microcontroller, we might only have a few hundred bytes of ROM, and a similar space for RAM. The device on the left-hand side is an MC6811, and only has 8 KB of ROM and 256 bytes of on-chip RAM. We thus need efficient code for our encryption, as there needs to be space for the main application software, too.

One method which focuses on creating an extremely small footprint is the PHOTON-Beetle method \cite{chakraborti2018beetle, PHOTON, asecuritysite_19090}. Overall it is a lightweight block cipher and was written by Zhenzhen Bao, Avik Chakraborti, Nilanjan Datta, Jian Guo, Mridul Nandi, Thomas Peyrin, and Kan Yasuda. It uses the sponge-based mode Beetle with the P256 for the permutation and supports both authenticated encryption (AE) and hashing. PHOTON-Beetle AEAD and PHOTON-Beetle hashing are finalists for NIST's competition on lightweight cryptography.

The Beetle family of cryptography methods integrates a lightweight, sponge-based authenticated encryption. When this is linked with the PHOTON permutation (PHOTON-256), it achieves an extremely small footprint. In tests, a 64-bit security version of PHOTON-Beetle consumes less than 600 LUTs (LookUp Tables) on an FPGA, compared with 1,000 LUTs for COFB-AES (COmbined FeedBack-AES).

PHOTON-Beetle can be optimized for either a low ROM environment (where the code needs to be compact) or for speed. For PHOTON-Beetle AEAD, on 8-bit microcontrollers with low ROM sizes, the ROM code size is less than 2,200 bytes, and adding a hashing method on top of this, only adds another 300 bytes of ROM. The requirement for memory, too, is small and where it only requires 100 bytes of RAM. The average speed is around 8,200 cycles per byte for encryption. For the PHOTO-Beetle AEAD mode which focuses on speed, the ROM code size is less than 4,100 bytes, with hashing adding 300 bytes. The average speed is around 4,900 cycles per byte for encryption.
\subsection{Romulus}
Romulus is a lightweight block cipher and was written by Tetsu Iwata, Mustafa Khairallah, Kazuhiko Minematsu and Thomas Peyrin \cite{iwata2020duel, Rom, asecuritysite_36089}. The NIST competition for lightweight cryptography has reached the final stage, and with a shortlist of 10 candidates. Each differs in their approach, but they aim to create a cryptography method that is secure, has a low footprint, and is robust against attacks. So while many of the contenders, such as ASCON, GIFT and Isap, use the sponge method derived from the SHA-3 standard (Keccak), Romulus takes a more traditional approach and looks towards a more traditional lightweight crypto approach. Overall it is defined as a tweakable block cipher (TBC) and which supports authenticated encryption with associated data (AEAD). For its more traditional approach, it uses the SKINNY lightweight tweakable block cipher. 

SKINNY is a light-weight block cipher. It has a 64-bit or 128-bit block size, and a key size of 64 bits, 128 bits and 256 bits. The methods are SKINNY-64-64 (64-bit block, 64-bit key and 32 rounds); SKINNY-64-128 (64-bit block, 128-bit key, and 36 rounds); SKINNY-64-192 (64-bit block, 192-bit key, and 40 rounds); SKINNY-128-128 (128-bit block, 128-bit key, and 40 rounds); SKINNY-128-256 (128-bit block, 256-bit key, and 48 rounds); SKINNY-128-384 (128-bit block, 384 key, and 56 rounds). For a 64-bit block it uses a 4x4 matrix for nibbles, and a 4x4 matrix of bytes for a 128-bit block size. :

For a 64-bit block it uses a 4x4 matrix for nibbles (4 bits), and a 4x4 matrix of bytes for a 128-bit block size. For the 4x4 matrix, each round we have operations of SC (SubCells); AC (AddConstants); ShiftRows (SR); and MixColums (MC):

For the SubCells (SC) we either have a 4-bit S-box (for 64-bit block) or a 8-bit S-box (for 128-bit block):

The AddRoundTweakey (ART) process takes part of the key, and applies it within each round.

\subsection{Sparkle}

Sparkle is a family of permutations. Schwaemm \cite{beierle2019schwaemm} is a light-weight cryptography method and provides confidentiality, integrity and authentication, while Esch provides a hashing method that is preimage and collision resistant. Both methods use a sponge construction using a cryptographic permutation (as used in SHA-3). Esch256 (Efficient, Sponge-based, and Cheap Hashing) implements the hashing method for a 256-bit hash, and has a block size of 16 bytes, a security level of 128 bits and a data limit of up to 2132. It was designed by Christof Beierle, Alex Biryukov, Luan Cardoso dos Santos, Johann Großschädl, Léo Perrin, Aleksei Udovenko, Vesselin Velichkov, and Qingju Wang. Schwaemm stands for Sponge-based Cipher for Hardened but Weightless Authenticated Encryption on Many Microcontrollers, and is also the Luxembourgish word of "sponge". 

The submitted version for the NIST competition includes Schwaemm \cite{beierle2019schwaemm, asecuritysite_15221} which is a lightweight cryptography method and provides confidentiality, integrity and authentication and Esch which provides a hashing method that is preimage and collision-resistant. Both methods use a sponge construction using a cryptographic permutation (as used in SHA-3). Esch256 (Efficient, Sponge-based, and Cheap Hashing) implements the hashing method for a 256-bit hash and has a block size of 16 bytes, a security level of 128 bits and a data limit of up to $2^{132}$.

For Schwaemm, the name is derived from "Sponge-based Cipher for Hardened but Weightless Authenticated Encryption on Many Microcontrollers", and which is also the Luxembourgish word of "sponge". Esch256 is also a part of a name of a place in Luxembourg. SPARKLE is similar to SPARKX and its name derives from SPARx, but Key LEss.

For AEAD, we have Schwaemm128–128, Schwaemm256–128, Schwaemm192–192, and Schwaemm256–256, and which support block sizes of 16 (128 bits), 32 (256 bits), 24 (192 bits) and 64 bytes, respectively. These give a security level that ranges from 120 bits to 248 bits.

\subsection{TinyJambu}
TinyJAMBU defines a family of lightweight cryptography methods and was designed by Hongjun Wu and Tao Huang \cite{wutinyjambu}.

\subsection{Xoodyak}
Xoodyak comes from the Keccak research team \cite{daemen2020xoodyak, asecuritysite_96729}, and which was successful in the SHA-3 competition. Overall, Keccak was evaluated as the most efficient and secure hashing method.

Joan Daemon also co-authored the Rijndael cipher that eventually became AES. With Xoodoo permutation we can apply it with the Xoodyak function. With this, we store a 384-bit state for the encryption and which relates to the sequence of the input data. With this, we can create a fixed-length hash, a pseudo-random bit value, or an output of a variable length. This can thus produce either a hash function, a random bit stream, or an encryption method.
\section{Results}\label{sec:Background}

And, so, since 2016, researchers have been probing the submitted methods, and in 2022 NIST published the final 10: ASCON, Elephant, GIFT-COFB, Grain128-AEAD, ISAP, Photon-Beetle, Romulus, Sparkle, TinyJambu, and Xoodyak. A particular focus is on the security of the methods, along with their performance on low-cost FPGAs/embedded processes and their robustness against side-channel attacks.

The current set of benchmarks includes running on an Arduino Uno R3 (AVR ARmega 328P), Arduino Nano Every (AVR ARmega 4809), Arduino MKR Zero (ARM Cortex M10+) and Arduino Nano 33 BLE (ARM Cortex M4F). These are just 8-bit processors and fit into an Arduino board. Along with their processing limitations, they are also limited in their memory footprint (to run code and also to store it). The lightweight cryptography method must thus overcome these limitations, and still, be secure and provide a good performance level. Running AES in block modes on these devices is often not possible, as there is not enough resources. Overall we use a benchmark for encryption — with AEAD (Authenticated Encryption with Additional Data) and for hashing. With AEAD we add extra information — such as the session ID — into the encryption process. This type of method can bind the encryption to a specific stream.

\subsection{ARM Cortex M3}

In Table \ref{tab:table01} [1], we see a sample run using an Arduino Due with an ARM Cortex M3 running at 84MHz. The tests are taken in comparison with the ChaCha20 stream cipher and defined for AEAD, and where the higher the value the better the performance. We can see that Sparkle, Xoodyak and ASCON are the fastest of all. Sparkle has a 100\% improvement, and Xoodyak gives a 60\% increase in speed over ChaCha20. Elephant, ISAP and PHOTON-Beetle have the worst performance for encryption (with around 1/20th of the speed of ChaCha20).

\begin{table*}
\caption{\label{tab:table01} Arduino Due with an ARM Cortex M3 running at 84MHz for encryption against ChaCha20 \cite{light01}}
\centering
\begin{tabular}{|l|l|l|l|l|l|l|l|l|}
\hline
Algorithm&Key Bits&Nonce Bits&Tag Bits&Encrypt 128~B&Decrypt 128~B&Encrypt 16~B &Decrypt 16~B&Aver
\\ \hline \hline
Schwaemm128-128 (SPARKLE)	&128	&128&	128	&1.6	&1.58	&2.84	&2.39	&2.01\\
Xoodyak 	&128	&128	&128	&1.66	&1.51	&1.73	&1.6	&1.62\\
ASCON-128	&128	&128&	128	&1.54	&1.44	&1.78	&1.68	&1.61\\
TinyJAMBU-128 	&128	&96	&64	&0.93	&0.95	&1.63	&1.61	&1.21\\
GIFT-COFB	&128	&128	&128	&1.01	&1.01	&1.16	&1.15	&1.08\\
Grain-128AEAD	&128	&96	&64	&0.26	&0.26	&0.56	&0.56	&0.37\\
Romulus-M1	&128	&128	&128	&0.1	&0.11	&0.15	&0.16	&0.13\\
PHOTON-Beetle-AEAD-ENC-128	&128	&128	&128	&0.06	&0.07	&0.11	&0.12	&0.08\\
ISAP-A-128	&128	&128	&128	&0.08	&0.08	&0.03	&0.04	&0.05\\
Delirium (Elephant)	&128	&96	&128	&0.04	&0.05	&0.06	&0.07	&0.05\\
\hline
\end{tabular}
\end{table*}

Not all of the finalists can do hash functions. Table \ref{tab:table02} outlines these.

\begin{table*}
\caption{\label{tab:table02} Arduino Due with an ARM Cortex M3 running at 84MHz for hashing against BLAKE2s \cite{NISTgov}}
\centering
\begin{tabular}{|l|l|l|l|l|l|}
\hline
Algorithm	& Hash Bits	& 1024 bytes	& 128 bytes	& 16 bytes	& Average\\
\hline\hline
Esch256 (SPARKLE) 	&256	&0.89	&0.78	&1.5	&1.06\\
Xoodyak 	&256	&0.71	&0.65	&1.43	&0.93\\
GIMLI-24-HASH	&256	&0.54	&0.47	&0.86	&0.62\\
ASCON-HASH 	&256	&0.51	&0.41	&0.63	&0.52\\
PHOTON-Beetle-HASH	&256	&0.01	&0.01	&0.05	&0.02\\
\hline
\end{tabular}
\end{table*}

Again, we see Sparkle and Xoodyak in the lead, with Sparkle actually faster in the test than BLAKE2s, and Xoodyak just a little bit slower. ASCON has a weaker performance, and PHOTON-Beetle is relatively slow. For all the tests, the ranking for authenticated encryption is (and where the higher the rank, the better):

14 SPARKLE
12 Xoodyak
12 ASCON
10 TinyJAMBU
9 GIFT-COFB, Gimli
4 Grain-128AEAD,KNOT
0 Elephant, ISAP, PHOTON-Beetle

and for hashing SPARKLE and Xoodyak are ranked the same:

7 SPARKLE, Xoodyak 5 Gimli 3 ASCON 0 PHOTON-Beetle

\subsection{Uno Nano performance}

For AEAD on Uno Nano Every [2], the benchmark is against AES GCM. We can see in \ref{tab:table03} , that SPARKLE is 4.7 times faster than AES GCM for 128-bit data sizes, and Xoodyak comes in second with a 3.3 times improvement over AES GCM. When it comes to 8-bit data sizes TinyJambu actually is the fastest, but where Sparkle and Xoodyak still perform well. PHOTON-Beetle, Grain128 and ISAP do not do well, and only slightly improve on AES GCM. In fact, Grain128 and ISAP are actually slower than AES GCM.

\begin{table*}
\caption{\label{tab:table03} Uno Nano for AEAD against AES GCM and showing cycles (showing fastest of the method)}
\centering
\begin{tabular}{|l|l|l|l|l|l|l|l|l|l|l|}
\hline
Algorithm&Impl.&Primary&Flag&Size&Enc(0:8)&Dec(0:8)&Enc(128:129)&Dec(128:128)&Bench.(128)&Bench.(8)
\\ \hline \hline
sparkle       &rhys	          &yes&	   O3	&12290	&1276	&1316	&4648    &5072  &4.7  &3.3\\
Xoodyak       &XKCP-AVR8	  &yes&    O3	&4560	&2596	&2608	&7184    &7128  &3.3  &1.6\\
knot	      &$avr8_speed$   &no&	   Os	&1664	&2124	&2140	&8144    &8160  &2.9  &2\\
ascon 	      &rhys	          &no&     O3	&5180	&1240	&1284	&8056    &8488  &2.8  &3.3\\
GIFT-COFB     &rhys	          &yes&    O1	&23312	&1852	&1892	&8220    &8776  &2.7  &2.2\\
saeaes	      &ref	          &no&     O3	&17062	&1208	&1212	&8992    &9004  &2.6  &3.4\\
hyena	      &rhys           &yes&    O3	&293860	&1912	&1964	&8960    &9396  &2.5  &2.2\\
elephant      &rhys           &no&     O3	&13106	&1924	&1948	&9260    &9796  &2.4  &2.2\\
estate	      &ref            &yes&    O3	&9434 	&1424	&1448	&10276   &10292 &2.3  &2.9\\
romulus	      &rhys           &no&     O3	&19346 	&1632	&1676	&10152   &10568 &2.2  &2.5\\
spook	      &rhys           &no&     O3	&12942 	&2984	&2968	&10272   &10708 &2.2  &1.4\\
tinyjambu     &rhys           &yes&    O3	&9174 	&1232	&1288	&10364   &10888 &2.2  &3.4\\
subterranean  &rhys           &yes&    Os	&6042 	&3372	&3460	&10288   &10944 &2.2  &1.2\\
orange        &rhys           &yes&    O3	&12140 	&2500	&2536	&11200   &11620 &2    &1.7\\
gimli         &rhys           &yes&    O3	&21272 	&1920	&1956	&11944   &12360 &1.9  &2.2\\
skinny        &rhys           &no&     O1	&12452 	&1604	&1644	&12960   &14372 &1.7  &2.6\\
photon-beetle & $avr8_speed$  &yes&    Os	&3536 	&2444	&2472	&20076   &20092 &1.2  &1.7\\
{\bf reference}&rhys          &yes&    O2	&7874 	&4152	&4156	&23812   &23764 &1    &1\\
grain128aead  &rhys           &yes&    O2	&9532 	&3992	&3980	&30396   &30124 &0.8  &1\\
isap          &rhys           &no&     O2	&3824 	&20212	&20256	&42936   &43372 &0.5  &0.2\\
\hline
\end{tabular}
\end{table*}

And so for AEAD  (performance) the ordering is

1. Sparkle
2. Xoodyak
3. Ascon
4. GIFT-COFB.
5. Elephant.
6. Romulus.
7. Tiny Jambu.
8. PHOTON-Beetle.
9. Grain128
10. ISAP.

For hashing on an Uno Nano Every, Table \ref{tab:table04} shows a similar performance level as to the ARM Cortex M3 assessment. In this case, the benchmark hash is SHA-256, and we can see that it takes Sparkle twice as many cycles for a 128-bit hash, and 2.9 times for Xoodyak. PHOTON-Beetle is way behind with a 128-bit hash and which is 17.4 times slower than SHA-256. That said, though, PHOTON-Beetle could be more focused on reducing power consumption rather than speed. GIMLI and SKINNY are included to show a comparison with well-designed methods in lightweight hashing. It can be seen that every method beats SKINNY, but only SPARKLE and Xoodyak beat GIMLI.

\begin{table*}
\caption{\label{tab:table04}  Uno Nano for hashing against SHA-256 and showing cycles (showing fastest of the method for hashing)}
\centering
\begin{tabular}{|l|l|l|l|l|l|l|l|l|l|l|}
\hline
Algorithm&Impl.&Primary&Flag&Size&h(8)&h(16)&h(32)&h(64)&h(128)&Benchmark
\\ \hline \hline
{\bf reference}&$nacl_ref$    &yes&    O3	&18774 	&768	&768	&772     &1364  &1968  &1\\
sparkle       &rhys	          &yes&	   O1	&7912	&1036	&1036	&1468    &2272  &3884  &2\\
Xoodyak       &XKCP-AVR8	  &yes&    O3	&2604	&1284	&1288	&1924    &3192  &5732  &2.9\\
gimli         &rhys           &yes&    O3	&19554 	&1284	&1920	&2544    &3804  &6312  &3.2\\
ascon 	      &rhys	          &yes&    O3	&2178	&2972	&3552	&4736    &7088  &11784 &6\\
drygascon     &rhys           &no&	   O3	&15500	&4604	&4600	&6540    &10360 &17912 &9.1\\
photon-beetle & $avr8_speed$  &yes&    O3	&2948 	&2372	&2364	&6940    &16084 &34172 &17.4\\
skinny        &rhys           &yes&    O2	&9784 	&7048	&10556	&13976   &20952 &34896 &17.7\\
\hline
\end{tabular}
\end{table*}

And so for hashing (performance) the ordering is:
\begin{enumerate}
    \item Sparkle.
    \item Xoodyak.
    \item Ascon
    \item PHOTON-Beetle. 
\end{enumerate}

\section{Conclusion}\label{sec:Conclusion and future work}

NIST announced ASCON as the chosen method that will be published as NIST’s lightweight
cryptography standard later in 2023. The performance benchmarks as stated in this article put Xoodyak and Sparkle look better in terms of performance. But, there are other assessments, such as security, energy footprint, and memory footprint that were taken into consideration from NIST when choosing the strongest and most efficient lightweight algorithm.

\bibliographystyle{IEEEtran}
\bibliography{template}


\end{document}